\documentstyle[11pt,paspconf]{article}

\input{psfig.sty}

\begin{document}

\title{Light Elements and Cosmic Rays in the Early 
Galaxy\altaffilmark{1}}

\author{R. Ramaty}
\affil{NASA/GSFC, Greenbelt, MD 20771}

\author{Benzion Kozlovsky}
\affil{School of Physics and Astronomy, Tel Aviv University, Israel}

\author{Richard E. Lingenfelter}
\affil{CASS, UCSD, LaJolla, CA 92093}

\altaffiltext{1}{In press, Proc. of the Space Telescope/Goddard 
High Resolution Spectrograph Symposium, Publ. Astron. Soc. 
Pacific, 1997} 

\begin{abstract} We derive constraints on the cosmic rays 
responsible for the Be and part of the B observed in stars 
formed in the early Galaxy: the cosmic rays cannot be 
accelerated from the ISM; their energy spectrum must be 
relatively hard (the bulk of the nuclear reactions should occur 
at $>$30 MeV/nucl); and only 10$^{49}$ erg/SNII in high 
metallicity, accelerated particle kinetic energy could suffice 
to produce the Be and B. The reverse SNII shock could 
accelerate the particles. \end{abstract}

\section{Introduction}

Observations of stars of various metallicities [$Z/Z_\odot$ 
$\equiv$ (Fe/H)/(Fe/H)$_\odot$] have shown that the abundance 
ratios Be/Fe and B/Fe do not vary appreciably over a broad 
range of $Z/Z_\odot$ (e.g. Duncan et al. 1996). For $Z/Z_\odot$ 
less than about 0.1, most of the Fe is thought (Truran \& 
Timmes 1994) to be produced in Type II supernovae (SNII). 
Limiting our considerations to this range of metallicities, the 
constancy of B/Fe and Be/Fe strongly suggest that B and Be are 
also produced by processes related to SNIIs. The accepted 
paradigm for the origin of most of the Be is production in 
spallation reactions of accelerated particles. B, in addition 
to such reactions, is also produced in neutrino interactions in 
SNIIs (Woosley \& Weaver 1995). The observed Be abundance as a 
function of metallicity can be approximated by ${\rm Be}/{\rm 
H} \simeq 5.2 \times10^{-11} Z/ Z_\odot$ (Duncan et al. 1996) 
which, with the solar Fe/H=3$\times$10$^{-5}$, implies that 
${\rm Be}/{\rm Fe} \simeq 1.7 \times 10^{-6}$, independent of 
$Z/Z_\odot$. This constant ratio greatly simplifies the 
treatment of the evolution and thereby allows us to clearly 
exhibit the strong constraints that the Be and B data place on 
cosmic ray acceleration in the early Galaxy. 

\section{Evolution}

We can simplify the equations of Galactic evolution by 
considering the total Be, B and Fe contents of the Galactic gas 
instead of the abundance ratios of these elements relative to 
H. Assuming that the removal rates of these trace elements from the 
gas due to incorporation into stars, as well as the return 
rates due to stellar mass loss are small relative to their 
production rates (assumptions whose consistency for 
$Z/Z_\odot<$0.1 we demonstrate below), the Be and Fe contents 
(measured in number of atoms) can be written as
\begin{eqnarray}
{\rm Be}(t) = \int_0^t Q_{\rm Be}(t') dt';~~~
{\rm Fe}(t) = \int_0^t Q_{\rm Fe}(t') dt'~~,
\end{eqnarray}
where the $Q$'s are production rates (measured in atoms/sec). 
An integral similar to that for Be can also be written for B, 
but for now we limit our discussion to Be because B could 
contain a significant non-cosmic ray contribution due to 
neutrinos. Using the integral for Fe, we can replace the time 
variable in the integral for Be by the iron content Fe. We 
obtain
\begin{eqnarray}
{ {\rm Be} \over {\rm Fe}}({\rm Fe}) = {1 \over {\rm Fe} } 
\int_o^{\rm Fe} {Q_{\rm Be}(Fe') \over Q_{\rm Fe}(Fe') } d(Fe')~~.
\end{eqnarray}
Since Be/Fe is observed to be constant as a function of 
$Z/Z_\odot$, it will also be constant as a function of Fe as 
long as long as $Z/Z_\odot$ is a monotonic function of Fe. In 
this case the solution of Eq.~(2) is $Q_{\rm 
Be}(Z/Z_\odot)/Q_{\rm Fe}(Z/Z_\odot)$ = const. The calculated 
(Woosley \& Weaver 1995) $^{56}$Fe yields per SNII, averaged 
over the Scalo (1986) IMF, lead to a relatively constant 
$^{56}$Fe yield as a function of $Z/Z_\odot$ (ranging from 0.09 
to 0.15 M$_\odot$ per SNII for $Z/Z_\odot$ varying from 
10$^{-4}$ to 0.1). Taking the $^{56}$Fe yield per SNII equal to 
0.1 M$_\odot$ (consistent with SN87A observations), we obtain 
$Q_{\rm Be}$$\simeq$3.6$\times$10$^{48}$ atoms/SNII or 
2.7$\times$10$^{-8}$ M$_\odot$/SNII, independent of $Z/Z_\odot$.

To see when the removal of Be and Fe can be ignored, let 
$\dot M_g(t)$ be the removal rate of Galactic gas (mostly H) 
due to star formation. The rates $R(t)$ of Be and Fe removal 
(measured in g/sec) are then 9(Be/H)$\dot M_g(t)$ and 
56(Fe/H)$\dot M_g(t)$, respectively. Assuming that all stars of 
mass $>$10M$_\odot$ form Type II supernovae, the SNII formation 
rate is
\begin{eqnarray} 
\dot N_{\rm SNII} = {\dot M_g \over {\rm M}_\odot} {\int_{10}^\infty 
\phi(m) dm \over \int_{0.1}^\infty m \phi(m) dm } \simeq 0.005 
{\dot M_g \over {\rm M}_\odot} \end{eqnarray} 
where $\phi(m)$ is the Scalo IMF (measured in number of stars 
per unit stellar mass). With Be and Fe productions per SNII of 
$2.7\times 10^{-8}$ and 0.1 M$_\odot$, respectively, the Be and 
Fe production rates $P$ (measured in g/sec) are $\simeq$1.35$\times 
10^{-10} \dot M_g$ and $\simeq$5$\times 10^{-4} \dot M_g$. 
Since ${\rm Be}/{\rm H}$$\simeq$5.2$\times$10$^{-11}$$Z/Z_\odot$ 
and ${\rm Fe}/{\rm H}$$\simeq$3$\times$10$^{-5}$$Z/Z_\odot$, the 
production to removal ratio for both Be and Fe becomes 
\begin{eqnarray}
{P\over R} \simeq {0.3 \over (Z/Z_\odot)}. 
\end{eqnarray}
Thus, for $Z/Z_\odot$$<$0.1, $P/R$$>$3 (becoming $\simeq$300 
for $Z/Z_\odot$=10$^{-3}$), allowing us to ignore the removal. 
This important result is simply the consequence of the fact 
that, while the Be and Fe productions are independent of the 
metallicity of the ISM, their removal is proportional to the 
metallicity. 

We have shown (Ramaty et al. 1996) that the early Galactic B/Be 
data and the predicted B-to-Be production ratio are consistent 
with a non-negligible contribution of $^{11}$B production by 
neutrino spallation in SNIIs (Woosley \& Weaver 1995). As both 
the B-to-Be cosmic ray production ratio as well as the IMF 
averaged $^{11}$B production by neutrinos are nearly independent 
of metallicity, our results for Be apply also for B.

\section{Production}

Be and B production by cosmic rays depends on a variety of 
parameters, but the required constant production, $Q_{\rm 
Be}$$\simeq$3.6$\times$10$^{48}$ atoms/SNII independent of 
metallicity, sets strong constraints on these parameters. We 
have developed a LiBeB production code (Ramaty et al. 1996) in 
which a population of accelerated particles of given 
composition, energy spectrum and total power $\dot W$ is 
injected into an ambient medium of given composition. We have 
taken into account all primary and secondary reaction paths 
leading to LiBeB from CNO. We assume that the transport of the 
accelerated particles in the ISM is characterized by a single 
parameter, $X_{\rm esc}$, the particle escape path length from 
the Galaxy which we take to be energy independent. The code 
employs updated nuclear cross sections and works equally well 
at nonrelativistic and ultrarelativistic particle energies. For 
the ambient medium composition we use solar abundances scaled 
with metallicity. For the accelerated particles we use three 
compositions: 
\begin{enumerate} 
\item SSz, identical to the ambient composition; this would be 
the composition if the accelerated particles were accelerated 
directly out of the ISM.  \item SNII, ejecta of Type II 
supernovae for various $Z/Z_\odot$ (12-40 M$_\odot$ 
progenitors, Woosley \& Weaver 1995), averaged over the Scalo 
IMF; these compositions would be appropriate if the accelerated 
particles were accelerated directly from the ejecta of such 
supernovae. The SNII composition is very weakly dependent on 
metallicity; p/O$\simeq$100, $\alpha$/O$\simeq$20, 
C/O$\simeq$0.2. \item SNII$_{\rm metal}$, identical to SNII but 
with p/O=$\alpha$/O=0.
\end{enumerate} 
We employ a shock acceleration spectrum modified with a high 
energy turnover, 
$Q(E)$$\propto$($p^{-s}$/$\beta$)exp($-E/E_o$), were $Q$ is the 
injection rate of the accelerated particles, $p$ and $E$ are 
momentum and kinetic energy, both per nucleon, $c\beta$ is 
velocity, and $E_o$ is a parameter. The deposited power that 
accompanies the nuclear reactions is $\dot W$=$\sum_{i} A_i 
\int_0^{\infty} Q(E) dE$. Fig.~1 shows $Q_{\rm Be}/\dot W$ as a 
function of $Z/Z_\odot$ (panel a) and $E_o$ (panel b). The Be 
production per SNII is $\eta W_{\rm SNII}(Q_{\rm Be}/\dot W)$, 
where $W_{\rm SNII}$ is the mechanical energy of an SNII 
available for acceleration and $\eta$ is the fraction of this 
energy going into the particles that produce the Be. 

\section{Discussion}

For the SNII$_{\rm metal}$ composition, which contains no 
accelerated protons and $\alpha$ particles, the Be is produced 
only by accelerated CNO interacting with ambient H and He; thus 
$Q_{\rm Be}/\dot W$ is not dependent on the ambient medium 
metallicity (Fig.~1a). This is also true for SNII since the 
metallicity of this composition  is high enough so that up to 
$Z/Z_\odot$=0.1 the contributions of accelerated protons and 
$\alpha$'s is negligible. Thus, if $\eta W_{\rm SNII}$ is 
independent of $Z/Z_\odot$ (a reasonable assumption), then both 
SNII (for $Z/Z_\odot$$<$0.1) and SNII$_{\rm metal}$ (for all 
$Z/Z_\odot$) produce constant Be yields per SNII, as required 
by the data. On the other hand, for SSz, for which the protons 
and $\alpha$'s make a major contribution, the Be yield is 
increasing with metallicity (Fig.~1a), a result which is not 
compatible with the observed constant Be/Fe. This means that 
the accelerated particles responsible for the Be production 
should be accelerated from freshly nucleosynthesized matter 
before it mixes with the ambient ISM. That in the low 
metallicity era of our Galaxy accelerated CNO, and not protons 
and $\alpha$'s, were responsible for the production of the Be 
and B was pointed out by Duncan, Lambert \& Lemke (1992) and 
Cass\'e, Lehoucq, \& Vangioni-Flam (1995). If $W_{\rm 
SNII}$=10$^{51}$ erg, we require that $Q_{\rm Be}/\dot 
W$$>$1.2$\times$10$^{-2}$ to produce 
3.6$\times$10$^{48}$ Be atoms per SNII with an acceleration 
efficiency $\eta$$<$0.3. This constraint provides another 
argument against SSz for which $Q_{\rm Be}/\dot 
W$$<$2$\times$10$^{-3}$ when $Z/Z_\odot$$<$0.1 (Fig.~1a). For 
SNII, $Q_{\rm Be}/\dot W$$>$1.2$\times$10$^{-2}$ implies that 
$E_o$$>$50 MeV/nucl (Fig.~1b), i.e. the accelerated particle 
spectrum must be relatively hard. The energetic constraint on 
$E_o$ is less severe for the SNII$_{\rm metal}$ composition.

From Fig.~1b the minimum energy ($\eta$$W_{\rm 
SNII}$$\simeq$10$^{49}$ erg) required to produce the Be is 
obtained for the SNII$_{\rm metal}$ composition. As this value 
is much lower than the energy per supernova imparted to the 
Galactic cosmic rays (a few times 10$^{50}$ ergs), the Be could 
be produced with less energy than needed to maintain the cosmic 
rays. The particles could be accelerated by the reverse 
supernova shock in all SNIIs, a scenario which could lead to 
the preferential acceleration of metals. Thus as little as 3\% 
of the total energy in accelerated particles could go into 
accelerating high metallicity SNII ejecta to produce the bulk 
of the Be and part of the B (spallation by neutrinos producing 
the rest). The remaining 97\% would go into accelerating cosmic 
rays out of the low metallicity ISM. These cosmic rays produce 
very little Be and B for Z/Z$_\odot$ $<$0.1.

\begin{figure}
\psfig{figure=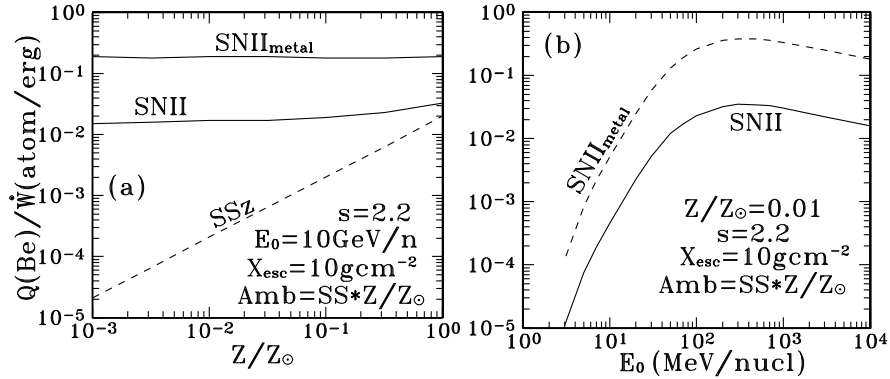,height=2in}
\caption{Be production rate per deposited cosmic 
ray power.} \label{fig-1}
\end{figure}

\end{document}